# Origin of the extremely high elasticity of bulk emulsions, stabilized by *Yucca Schidigera* saponins

Sonya TSIBRANSKA,[1] Slavka TCHOLAKOVA,[1*] Konstantin GOLEMANOV,[1] Nikolai DENKOV,[1] Luben ARNAUDOV,[2] Eddie PELAN,[2] Simeon D. STOYANOV[2,3,4]

[1]*Department of Chemical and Pharmaceutical Engineering*
*Faculty of Chemistry and Pharmacy, Sofia University*
*1 J. Bourchier Ave., 1164 Sofia, Bulgaria*

[2] *Unilever R&D, Vlaardingen, The Netherlands*

[3]*Laboratory of Physical Chemistry and Colloid Science, Wageningen University, 6703 HB Wageningen, The Netherlands*

[4]*Department of Mechanical Engineering, University College London, Torrington Place, London WC1E 7JE, UK*

*Corresponding author:
Prof. Slavka Tcholakova
Department of Chemical and Pharmaceutical Engineering
Faculty of Chemistry and Pharmacy, Sofia University
1 James Bourchier Ave., 1164 Sofia
Bulgaria

Phone: (+359-2) 962 5310
Fax:    (+359-2) 962 5643
E-mail: SC@LCPE.UNI-SOFIA.BG

# Abstract

We found experimentally that the elasticity of sunflower oil-in-water emulsions (SFO-in-W) stabilized by *Yucca Schidigera Roezl* saponin extract, is by > 50 times higher as compared to the elasticity of common emulsions. We revealed that strong specific interactions between the phytosterols from the non-purified oil and the saponins from the *Yucca* extract lead to the formation of nanostructured adsorption layers which are responsible for the very high elasticity of the oil-water interface and of the respective bulk emulsions. Remarkably, this extra high emulsion elasticity inhibits the emulsion syneresis even at 65 vol % of the oil drops – these emulsions remain homogeneous and stable even after 30 days of shelf-storage. These results demonstrate that the combination of saponin and phytosterols is a powerful new approach to structure oil-in-water emulsions with potential applications for formulating healthier functional food.

# 1. Introduction.

Extract from Yucca schidigera is commercial source of steroid saponins and it is used in beverage and food industries (Cheeke, 2000) with the approval of U.S. Food and Drug Administration (FDA – 2014). The chemical composition of Yucca Schidigera extract is widely studied in the literature (Ralla, Salminen, Tuosto, & Weiss, 2017; Kowalczyk, Pecio, Stochmal, & Oleszek, 2011; Miyakoshi et al., 2000; Oleszek et al., 2001; Piacente, Pizza, & Oleszek, 2005) and it contains around 9.5 % saponins; 1.4 % proteins; 22.5 % sugars; 1.2 % minerals (Ralla, Salminen, Tuosto, & Weiss, 2017) and some phenols (Piacente, Pizza, & Oleszek, 2005). The combination of saponins and phenols in Yucca extracts leads to their strong biological effects, such as growth-inhibitory activity against different yeasts (Miyakoshi et al., 2000); antibacterial and antiprotozoal effects against *Butyrivibrio fibrisolvens* and *S. ruminantium* (Wallace, Arthaud, & Newbold, 1994), antioxidant effect (Piacente, Montoro, Oleszek, & Pizza, 2004), antiplatelet effect (Olas, Wachowicz, Stochmal, & Oleszek, 2002); and antiinflamatory activity (Marzocco et al., 2004; Piacente, Pizza, & Oleszek, 2005).

Along with the very strong biological activity, the Yucca extracts show also significant surface activity (Oleszek & Hamed, 2010; Golemanov, Tcholakova, Denkov, Pelan, & Stoyanov, 2012; Pagureva et al., 2016). The adsorption layers of *Yucca* saponins exhibit purely viscous rheological response, even at the lowest shear stress applied (Golemanov, Tcholakova, Denkov, Pelan, & Stoyanov, 2012), whereas the adsorption layers of the triterpenoid *Quillaja* saponins behave like a viscoelastic two-dimensional body. The difference in the properties of Yucca and Quillaja adsorption layers was explained with the different molecular structure of the aglycones of these saponins – steroid vs. triterpenoid (Golemanov, Tcholakova, Denkov, Pelan, & Stoyanov, 2012; Pagureva et al., 2016).

Yucca extracts are studied also as foaming agent in different applications (Sucharzewska, Stochmal, & Oleszek, 2003; Park, Plahar, Hung, McWatters, & Eun, 2005). It was found that Yucca extract had beneficial influence on the structure (dry matter content, hardness, density and porosity) of sugar-candy foam products, both in newly prepared and in candies stored over two months (Sucharzewska, Stochmal, & Oleszek, 2003). This extract affects positively the properties of deep-fat-fried cowpea paste called "Akara" by increasing the foaming capacity of the paste (Park, Plahar, Hung, McWatters, & Eun, 2005). Tcholakova and co-workers (Tcholakova et al., 2017) showed that the rate of Ostwald ripening in Yucca stabilized foams is faster as compared to Quillaja saponins which was explained with the lower surface elasticity of adsorption layers formed from Yucca saponins.

The emulsifying properties of Yucca extracts were studied by Ralla et. al (Ralla, Salminen, Tuosto, & Weiss, 2017). They showed that Yucca extract can be used to prepare 500 nm droplets from 10 vol % medium chain triglyceride oil Miglyol 812N using 1 wt % saponin and high pressure homogenizer (Ralla, Salminen, Tuosto, & Weiss, 2017). The emulsions were stable at pH between 5 and 9 and showed only minor increase in droplet size when heated up to 90 °C (Ralla, Salminen, Tuosto, & Weiss, 2017). The obtained results from Ralla et al. indicate that Yucca extracts can be used to stabilize emulsions.

In our previous study (Tsibranska et al., 2020) we investigated the effect of four triterpenoid saponins (Quillaja saponin; Escin; Berry saponin concentrate and Tea Saponin) on emulsion rheological properties. We found that the dilatational interfacial elasticity has very significant impact on the emulsion shear elasticity, intermediate effect on the dynamic yield stress, and no effect on the viscous stress of the concentrated emulsions.

Here we study the elasticity of emulsions of sunflower oil (SFO), stabilized by steroid saponins extracted from the plant *Yucca Schidigera Roezl*. We found that these emulsions exhibited extremely high elastic modulus, $G` \approx 2000$ Pa, at relatively low oil fraction, $\Phi_{OIL} = 65$ wt. %. We found that this unusually high emulsion elasticity comes from the synergistic action of two different factors: high interfacial elasticity and strong drop-drop adhesion in the bulk emulsions. Previous results published in the literature show that the adhesion between the emulsion drops can significantly increase the elastic modulus of the bulk emulsions (Alben, Holtze, Tadros, & Schurtenberger, 2012; Becu, Manneville, & Colin, 2006). However, only the combination of these two factors could explain the extremely high emulsion elasticity at drop fractions as low as 65 wt %, measured in our experiments. Furthermore, we revealed the high interfacial elasticity is created by a strong specific interaction between the saponins from the *Yucca* extract and the phytosterols present in the non-purified sunflower oil. To find the mechanistic explanations of the observed effects, we measured the interfacial modulus of the adsorption layers of *Yucca Schidigera* saponins on the oil-water interface with purified and non-purified oil and observed by optical microscopy the emulsion films of type oil-water-oil.

The article is organized as follows. The materials and methods are described in section 2. In section 3 we present the experimental results and their discussion. The main conclusions are summarized in section 4.

**2. Materials and methods.**

**2.1. Materials.**

We studied 3 saponin extracts of the plant species *Yucca Schidigera Roezl.*, products of Desert King, Chile: two liquid extracts with trade names Foamation P (FP) and Foamation 50 (F50), and one powder extract Foamation Dry 50 (FD). All these extracts contain around 9 % saponins according to their producer and to our analysis. For FD the saponin content is 9 wt %, for FP it is 10 wt %, and for F50 it is 9.5 wt %. The liquid extracts FP and F50 contain also 0.1 wt. % of sodium benzoate, added as preservative. The difference between FP and F50 is that FP is partially purified and contains lower concentration of admixtures (particulates and water-soluble components) as compared to F50. Powder extract FD contains also ca. 50 wt. % maltodextrin. In our experiments we studied the properties of aqueous solutions containing 1 wt % saponin, with added 10 mM NaCl as neutral electrolyte and 0.1 g/L $NaN_3$ as an antibacterial agent.

In comparative experiments we prepared also emulsions stabilized by two synthetic surfactants: the anionic sodium dodecyl sulfate (abbreviated as SDS, product of Acros Organics, cat. No: 23042-500) and the nonionic polyoxyethelene-8 tridecyl ether $C_{13}EO_8$ (abbreviated as ROX, product of Rhodia).

The sunflower oil was provided by Unilever R&D, The Netherlands. Sunflower oil consists of triglycerides of fatty acids (linoleic 48-74 %, oleic 10-14 %, palmitic 4-9 %, stearic 1-7 %) and contains numerous admixtures - lecithin, torcophenols, cartenoids and waxes. In the experiments we used purified and non-purified oils for comparison. The oil was purified from surface-active impurities by passing it through a glass column filled with chromatographic adsorbent Florisil (Supelco, Cat. No. 20280-U, PR60/100 Mesh) and silica gel 60 (Merck, Cat. No. 107734.2500) using the procedure of Gaonkar & Borwankar, 1991.

To test the purity of the oil, we measured the interfacial tension at the interface between oil and pure water. The interfacial tension of the purified SFO-water interface was 30.5 ± 0.5 mN/m at 20 °C. There was no significant decrease of this tension over time, up to 15 min, which indicated that there were no surface active substances remaining in the oil after its purification.

**2.2. Measurement of the interfacial tension of the solutions.**

To measure the interfacial tension (IFT) on the oil-water surfaces we used drop shape analysis (DSA) method on instrument DSA100R and software DSA1 (Krüss GmbH, Germany). The method consists of formation of a deformed drop from one of the phases in the other phase. The software automatically detected the profile of the drop and fitted it with the Laplace equation of capillarity to determine the drop volume, the drop surface area, the interfacial tension, $\sigma$, and the error of the best fit to the drop profile. We formed water drop on the tip of a glass capillary immersed in the oil phase. The measurements were performed at $T = 25$ °C.

**2.3. Measurement of the surface dilatational modulus.**

The surface dilatational modulus was measured on DSA100 instrument, equipped with Oscillating Drop Module (ODM module DS3260 by Krüss GmbH). Sinusoidal oscillations of the area of pendant water drops were applied and the changes in the area and in the interfacial tension of the drops were measured by using drop shape analysis (Möbius & Miller, 1998; Russev et al., 2008; Alexandrov, Marinova, Danov, & Ivanov, 2009). We performed experiments at 2, 5 and 10 s period of oscillation, corresponding to 0.5, 0.2 and 0.1 Hz oscillation frequency. The amplitude of deformation was varied in the range 0.1 to 5 %. Before each experiment, the layer was left to equilibrate for 15 minutes.

**2.4. Procedure for emulsion preparation.**

All emulsions were produced at room temperature (≈ 25 °C). We prepared an oil-in-water premix in Kenwood mixer by the following procedure. First, we added the aqueous saponin solution in the container, after that we added slowly the sunflower oil for a period of 4 min at speed of 1.5 rps (this speed is designated as level 3 on the dial of the instrument, controlling the rotation rate). After adding the oil, we homogenized this oil-water premix for additional 4 min at the same speed, 1.5 rps. After that we transferred this premix into the colloid mill (Magic Lab; MK; U078310) which is used to prepare the final emulsion. The gap in the colloid mill was set to be 398 μm and the rate of rotation was 10 000 rpm. The homogenizer was connected to a pump (ISMATEC; MCP-CPF Process IP65; Pump head - FMI212/QP.Q2.CSC/9004) which facilitated the flow of the emulsion. The pump operated at 500 rpm rotation rate. The total volume of the emulsion was fixed at 200 mL. In all experiments, the saponin concentration in the aqueous phase was 1 wt. %. The concentration of SDS and ROX was 4 and 5 wt. %, respectively. Higher concentrations of SDS and ROX are used because the emulsions at high oil volume fractions were unstable when SDS or ROX were used as emulsifiers at 1 wt %.

**2.5. Microscope observations of the emulsions.**

We observed the shape and the aggregation of the emulsion drops in the studied emulsions in transmitted light using optical microscope Axioplan (Zeiss, Germany), equipped with objective Zeiss Epiplan 50×. The emulsion samples were diluted with 1wt% saponin solution to observe the shape of the drops or with 2.5 mM SDS solution in the case of drop size measurements (the drops acquire spherical shape in SDS solutions). Afterwards, the diluted samples were placed in a borosilicate glass capillary with rectangular cross-section.

The drop size distribution was determined with microscope Axio Imager M2m (Zeiss, Germany). The same rectangle capillary with emulsion sample was observed in transmitted light. The microscope was equipped with a sample holder, which allows automatic scan of the samples, during which the microscope takes a series of pictures of the emulsion drops. The software of the microscope (Axio Vision) processes the recorded images, determines the radius of each drop, and calculates the mean surface-to-volume drop radius, $R_{32}$:

$$R_{32} = \sum_i R_i^3 \Big/ \sum_i R_i^2 \tag{1}$$

where the sum is taken over all measured drops in a given sample. For each sample we measured the radii of at least ≈ 10 000 drops.

**2.6. Measurements of the creaming kinetics of the emulsions.**

The dynamics of creaming of diluted emulsions is affected by the drop-drop interactions (e.g. by depletion forces or bridging interaction). To study this phenomenon we used the following protocol. First, we produced emulsions with $\Phi_{SFO}$ = 75 wt. % via the standard procedure, described in section 2.4. Then, this emulsion was placed in a Kenwood bowl and diluted down to $\Phi_{SFO}$ = 9 wt. %. The diluted emulsion was mixed with the Kenwood mixer for 40 s at 1.5 rps rotation, and for additional 40 s at 0.5 rps. After this mixing, we placed samples from the emulsion into two identical cylinders, with volume of 100 mL each. Afterwards we recorded the volumes of the separated serum and of the cream as a function of time. We observed a very clear boundary between the emulsion which contained drops, and the serum separated below the emulsion as a result of drop creaming. We measured the ratio $h(t)/h_0$ with time, where $h(t)$ is the cream height, and $h_0$ is the total emulsion height. These experiments were performed with extracts of *Yucca Schidigera* (Foamation Dry 50), again with concentration of 1 wt. % saponin in the aqueous phase.

### 2.7. Characterization of the emulsion stability to syneresis.

We characterized the emulsion stability to syneresis via the following protocol. The emulsions were placed in a tube with graduated volume of 50 mL. The height of the tube was 11.5 cm and the diameter was 2.8 cm. Each tube was filled with 45 mL emulsion. We monitored the drainage of the aqueous phase for a period of up to 1 month. The volume of the solution drained at the bottom of the tube was measured. The tubes were stored at room temperature in a polystyrene box.

### 2.8. Characterization of the rheological properties of the emulsions.

The emulsion rheological properties were studied with rotational rheometer Bohlin Gemini (Malvern Instruments, UK) or Discovery Hybrid Rheometer DHR-3 (TA instruments, USA). We used parallel-plates geometry with 40 mm diameter of the upper plate. On both plates we glued sandpaper, grade P1500, to suppress the possible slip between the sample and the plate (the so-called "wall slip"). Afterwards the rheological response of the emulsions was characterized in oscillatory deformation applying two rheological tests: (1) Amplitude sweep under strain control. The samples were left to thermally equilibrate in the rheometer for 60 s at $T = 20$ °C. The gap between the plates was set at 2000 µm, and oscillatory deformation was applied. The strain was varied logarithmically in the range from 0.05 to 50 %. The storage and the loss moduli, $G`$ and $G``$, were measured, as a function of the amplitude of oscillations, at constant frequency of 1 Hz. The elasticity of the emulsions was characterized with the elastic modulus of the emulsion in the plateau region ($\gamma_A < 0.5$). (2) Frequency sweep under strain control. The samples were left to thermally equilibrate in the rheometer for 60 s at $T = 25$ °C. The gap was set at 2000 µm, and a pre-shear was applied at constant shear rate (30 s$^{-1}$) for 60 s. Afterwards, the gap was lowered down to 1500 µm, and an oscillatory deformation was applied. The frequency was varied logarithmically from 0.003 to 1 Hz. The storage and the loss moduli, $G`$ and $G``$, were measured, as a function of the frequency of oscillations, at constant amplitude of deformation ($\gamma_A \approx 0.6$ %).

### 2.9. Observation of thin emulsion films in capillary cell.

Emulsion films of sub-millimeter size were formed and observed in a capillary cell using the method of Scheludko-Exerowa (Scheludko, 1967). The emulsion films were formed from a biconcave drop, placed in a short capillary, by sucking out liquid through a side orifice in the capillary. The space around the capillary was filled with oil, thus forming aqueous emulsion film, sandwiched between the oil phases. The films were observed in reflected light via optical microscope Axioplan (Zeiss, Germany), equipped with a long-distance objective Zeiss Epiplan 20×/0.40. From the intensity of the reflected light we determined the thickness of the emulsion films.

### 2.10. Gas Chromatography analysis

We used gas chromatography (GC) to analyze sunflower oil and purified sunflower oil, and used phytosterols as external standards. Chloroform solutions of these samples were derivatized by mixing with N,O-Bis(trimethylsilyl)trifluoroacetamide (BSTFA), anhydrous pyridine and an internal standard (cetanol) for 1 h at 60 ºC and diluted with 2,2,4-trimethylpentane.

The GC analyses were performed on a TRACE GC apparatus (ThermoQuest, Italy) equipped with autosampler AS 2000, both being donated kindly by Unilever R&D Laboratory in Colworth, UK (SEAC division). We used a capillary column Quadrex, USA, with the following specification: 5 % phenyl methylpolysiloxane, 10 m length, I.D. 0.53 mm, 0.1 μm film thickness. Cold on-column injection was used, at a secondary cooling time of 0.3 min. The injection volume was 1 μL. The oven was programmed as follows: start at 110 ºC, hold 1 min, ramp 1 to 180 ºC at 10 ºC/min, ramp 2 to 350 ºC at 40 ºC/min, hold 3 min. The flame-ionization detector (FID) temperature was set to 350 ºC. The carrier gas was helium, set at a constant pressure mode (80 kPa). The detector gases were hydrogen and air, with nitrogen as make-up gas. The secondary cooling gas was nitrogen with a purity of 99.99 %. All other gases were of 99.999 % purity.

### 2.11. Freeze fracture SEM.

A droplet of the studied sample was placed on top of a rivet and plunge-frozen in melting ethane. The sample was cryo-planed using a cryo-ultramicrotome (Leica Ultracut UCT EM FCS) to obtain freshly prepared cross-section. Cryo-planing was done first by a glass knife and the last sections were made with diamond knife (Diatome histo cryo 8 mm) at −120 °C. The rivet was mounted onto a holder and transferred into a Gatan Alto 2500 preparation chamber. To reveal the microstructures under the planed surface, the temperature of the sample was increased for a short period to −90 °C to remove a thin surface layer of frozen water by sublimation. This yielded a 3D view on the planed sample. The sample was sputter coated with platinum (30 s) for a better SEM contrast and to prevent charging by the electron beam during observation. The sample was imaged using a Zeiss Auriga field-emission SEM at −125 °C and an accelerating voltage of 3 kV.

### 3. Experimental results and discussion.

### 3.1. Rheological properties of the emulsions.

Figure 1A presents illustrative results for the elastic, $G`$, and viscous, $G``$, moduli of FD-stabilized emulsion of non-purified sunflower oil, as a function of the strain amplitude, $\gamma_A$. The observed dependences of $G`$ and $G``$ on $\gamma_A$ are typical for concentrated emulsions and foams (Princen, 2001; Mason, Bibette, & Weitz, 1996). At low values of $\gamma_A$, in the so-called "plateau region" both moduli do not change significantly with the increase of $\gamma_A$. At a certain value of $\gamma_A$, $G`$ starts to decrease steadily, while $G``$ passes through a maximum. Throughout the article we characterize the elasticity of the bulk emulsions with the value of the elastic modulus in the plateau region, viz. at $\gamma_A < 0.2$ %. The respective value is simply referred to as $G`$ in the text.

The experimental results for $G`$ and $G``$, as a function of the oscillation frequency, are presented in Figure 1B. One sees that the studied emulsions have much higher elastic modulus when compared to the viscous modulus in the whole frequency range. The emulsion elasticity measured in amplitude sweep experiment at 1 Hz is ≈ 3700 Pa, whereas it decreases to ≈ 2000 Pa in the frequency sweep experiments, which is due to the applied pre-shear in the beginning of the frequency sweep experiments. This pre-shear perturbs the attractive forces between the drops. However, direct experimental checks showed that the emulsion elasticity recovers completely when the sample is left to relax for several hours after the pre-shear. Results from amplitude sweep experiments before pre-shear, after pre-shear and after different times of storing the sample in the rheometer without shearing it, are compared in Figure S1 in Supplementary information. One sees that the elastic modulus decreases after the pre-shear, but it increases again after storing the sample for 6 h without shearing it.

Figure 1C presents the results for the emulsion elastic modulus (in the plateau region), as a function of the oil weight fraction. The measured elasticity of the emulsions is remarkably high: ≈ 2000 Pa at $\Phi_{SFO} = 65$ wt. % and it increases further with the increase of oil fraction. Note that the emulsions typically have elasticity of ≈ 400 Pa, i.e. an order of magnitude lower, at $\Phi_{SFO} = 75$ vol. % (Tsibranska et al., 2020). Another reference worthy to note is the elasticity of hexadecane emulsions, stabilized by *Quillaja* saponin extract QD with very high interfacial elasticity, which was ≈ 800 Pa at 75 vol %, whereas the FD-stabilized emulsions have emulsion elasticity of 4000 Pa at the same $\Phi_{SFO}$. The difference between the FD and QD stabilized emulsions is even bigger at 65 wt % where the elasticity of the conventional emulsions approaches zero and even the QD-stabilized emulsions, with very high interfacial elasticity, exhibit bulk emulsion elasticity < 50 Pa.

These results demonstrate that the elasticity of FD-stabilized emulsions is unusually high when compared to other known emulsions. This exceptional elasticity has significant impact also on the stability to syneresis of the respective emulsions. Typically the time required for water drainage from 65 wt % SFO-in-water emulsions is < 1 day, whereas the emulsions prepared from FD extract possess very high stability to syneresis– no water separation is detected on the container bottom for more than 30 days of shelf-storage.

In our efforts to clarify the reason for this extremely high elasticity of the FD stabilized emulsions, we prepared emulsions with liquid saponin extracts which do not contain maltodextrins. As seen in Figure 1D, the elasticities of these emulsions are also very high which proves that the maltodextrins present in the dry extract FD are not the reason for the observed high bulk elasticity of the FD stabilized emulsions.

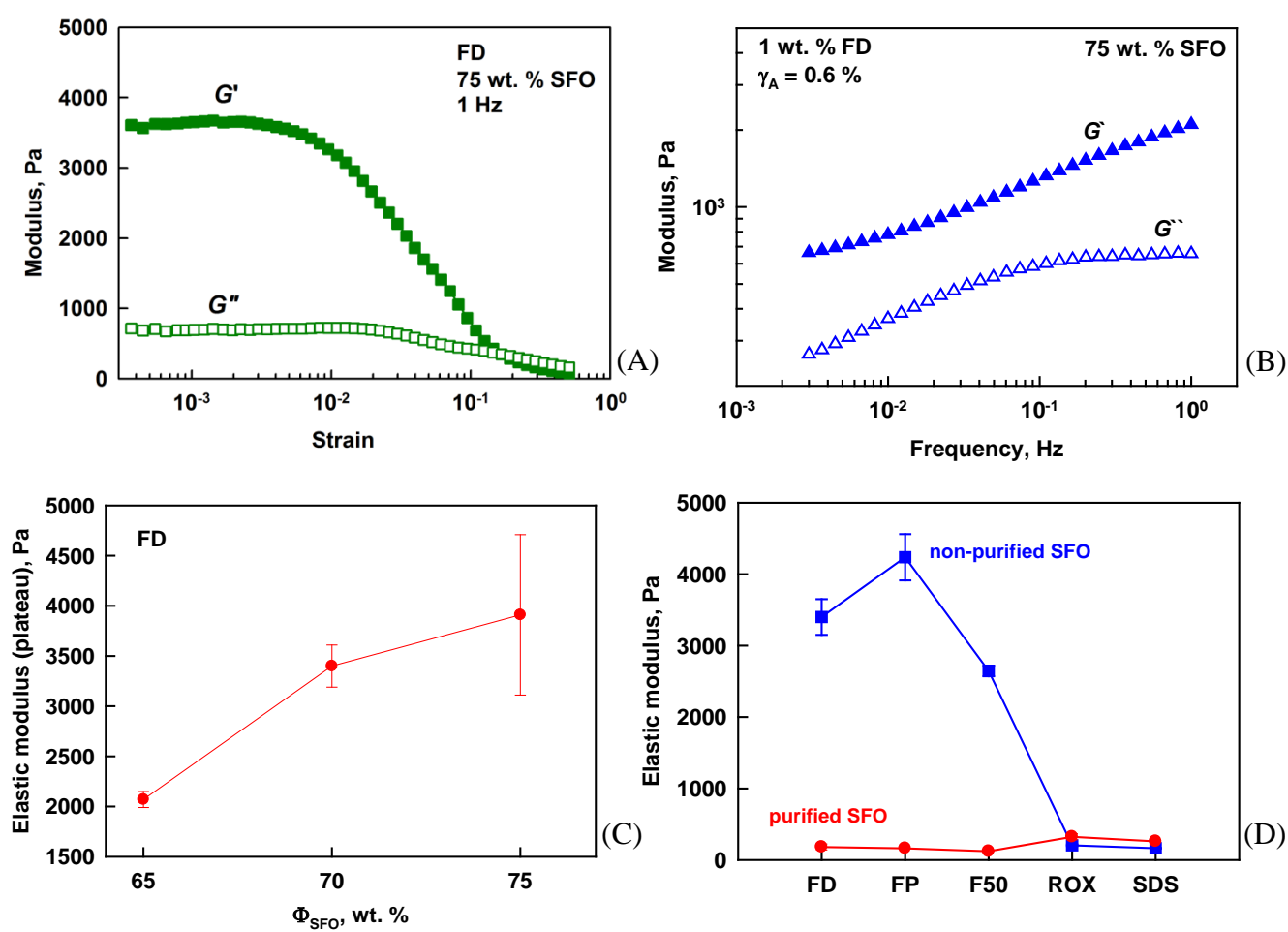

**Figure 1.** (A) Viscoelasticity of emulsions of sunflower oil-in-water, stabilized with FD. Elastic (full symbols) and viscous (empty symbols) moduli as functions of (A) amplitude of deformation at 1 Hz and (B) frequency of oscillation at 0.6 % strain for emulsion with $\Phi_{SFO}$ = 75 wt. %.; (C) Elastic modulus (plateau region) vs. oil volume fraction. (D) Elastic modulus (plateau region) of sunflower oil-in-water emulsions stabilized by different extracts of *Yucca Schidigera*, ROX or SDS. $\Phi_{SFO}$ = 75 wt. %. Concentrations: 1 wt. % saponin; 5 wt. % ROX; 4 wt. % SDS.

To check whether the measured high elasticity of the emulsions prepared with non-purified oil is related to some minor components present in the oil, we prepared also emulsions with purified SFO oil. These results are shown in Figure 1D and one sees that the bulk elasticities of these emulsions are by more than 10-times lower as compared to those of emulsions prepared with non-purified oils. Furthermore, the elasticities of the bulk emulsions prepared with purified SFO and FD saponins are similar to those of the emulsions stabilized by low-molecular-mass surfactants ROX and SDS. Note that the purification of SFO slightly increases the elasticity of ROX and SDS-stabilized emulsions, due to the higher interfacial tension measured with purified SFO.

These results reveal that the ultra-high elasticity of the FD-stabilized emulsions is related to the presence of some minor components, such as phytosterols or diglycerides that are present in the non-purified SFO. Apart from the changes in the rheological properties of the bulk emulsions, we observed also a significant decline in the emulsion stability when purified SFO was used for emulsion preparation. We shelf-stored the prepared emulsions at room temperature and observed their behavior. The emulsions with 75 wt% non-purified SFO were stable (without water or oil separation) for more than 12 months, whereas the emulsions prepared with purified SFO released top layer of oil within less than 6 months after their preparation, i.e. drop coalescence occurred in the emulsions prepared with purified SFO. In other words, the component(s) in the oily phase which affects strongly the rheological properties of the bulk emulsions suppresses also the drop-drop coalescence in these emulsions.

### 3.2. Composition of purified and non-purified sunflower oil

To clarify which components lead to this significant difference in the properties of the emulsions prepared by purified and non-purified SFO, we performed GC analysis of these two oils. Figure 2 compares GC chromatograms of SFO before and after its purification. The whole chromatogram is presented on Figure 2A, whereas the main differences between purified and non-purified SFO are enlarged in Figure 2B. Non-purified SFO contains much higher amounts of diglycerides (DG) and phytosterols. Therefore the possible substances which could be involved in the high emulsion elasticity are the diglycerides and phytosterols present in the non-purified oil which are missing in the purified oil.

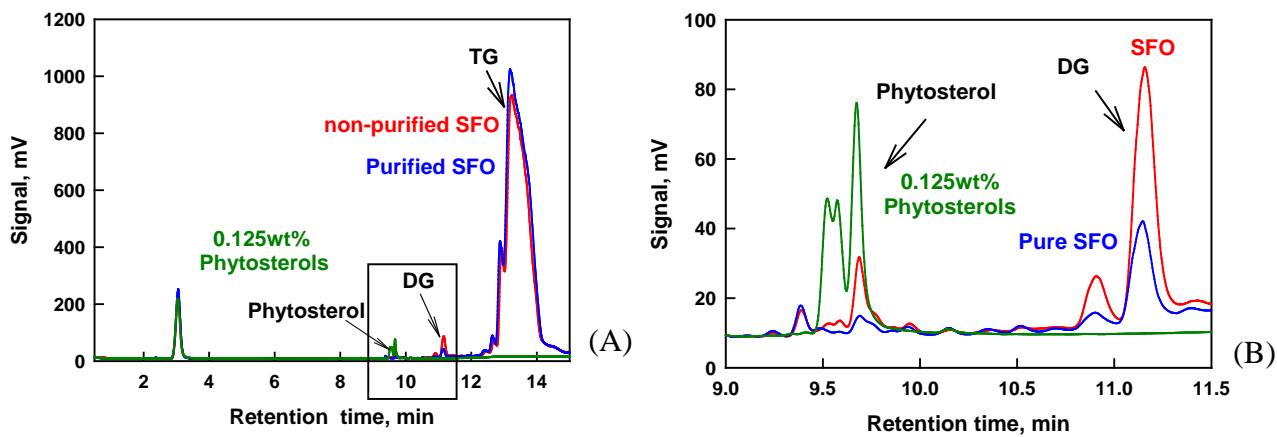

**Figure 2.** GC chromatogram of Sunflower oil before (red curve) and after purification (blue curve). In (A) the whole chromatogram is shown, whereas (B) shows the range of retention times at which significant differences between the two chromatograms are observed. For comparison, the peaks from reference phytosterols are also shown with green curve.

### 3.3. Role of oil purification for the surface and film properties of *Yucca* solutions

One should expect that the observed difference in the elasticities of the emulsions prepared with purified and non-purified oil is related to differences in the surface and emulsion film properties of these systems. To check this possibility, we measured the dilatational interfacial elasticity of the FD adsorption layers against purified and non-purified SFO and observed the respective emulsion films.

Figure 3 shows drops of 1wt% FD solution immersed in several types of oil. When non-purified SFO was used, we observed the formation of wrinkles on the oil-water interface when we shrink the drop surface. These wrinkles serves as clear prove for solidification of the adsorption layer which is related to extra-high interfacial elasticity, see Figure 3A. In contrast, when purified SFO was used, the interface was fluid, no wrinkles were seen, and the measured interfacial elasticity was very low, ca. 1 mN/m, see Figure S2 in Supporting information.

When we added DG to purified SFO oil, we found that the measured surface elasticities with purified SFO and with purified SFO+DG were very similar and low, see Figure S2 in Supporting information. This result shows that the DGs present in the non-purified oil do not interact with the Yucca saponins to create highly viscoelastic interfaces. In contrast, when we added 0.15 wt % phytosterols to purified SFO, we observed the formation of elastic skin, as in the case of non-purified oil, see Figure 3C. Thus we conclude that the skin formation on the surface of a drop in contact with non-purified SFO oil is related to a very strong specific interaction between the saponins in the *Yucca* extract and the phytosterols present in the non-purified oil.

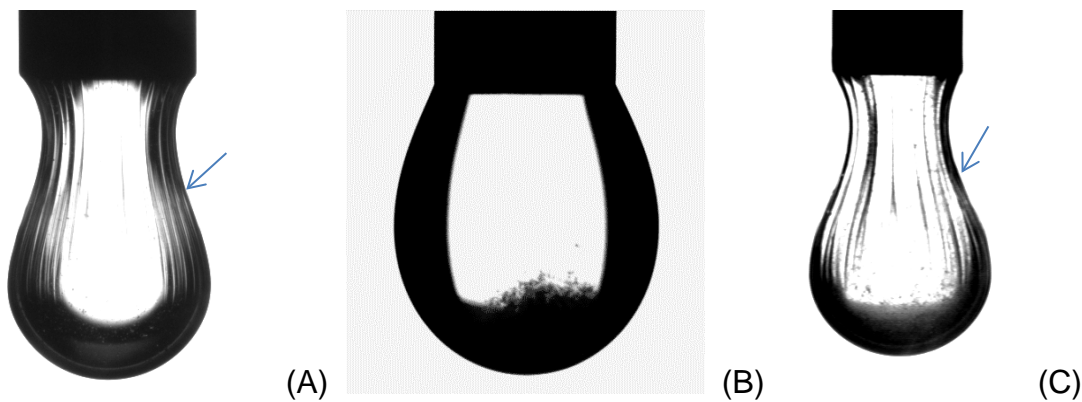

**Figure 3.** Drop of 1wt% solution of FD in sunflower oil. (A) non-purified SFO; (B) Purified SFO; (C) Purified SFO+0.15wt% Phytosterols. Wrinkles are formed on drop surface for (A) and (C) as indicated by arrows.

To check whether similar skins are formed on the surface of the oil drops in real emulsions, we used freeze fracture SEM to observe the surface of drops prepared with non-purified oil and stabilized by FD. Figure 4 shows three images at different magnifications. One sees peculiar nanopatterning on the surface of the oil drops. Most probably, this nano-structure is created by specific alignment of the saponin and phytosterol molecules in the mixed adsorption layers which leads to the formation of nano-sized ordered domains.

Interestingly, similar surface nanostructures were theoretically predicted in computer modelling of the molecular structure of adsorption layers of escin – another saponin which produces highly elastic adsorption layers (Tsibranska, Ivanova, Tcholakova, & Denkov, 2019). In the latter study we found that the escin molecules spontaneously self-assemble in their adsorption layers to form structured domains and surface nano-wrinkles. This structuring and the related extra-high elasticity of the escin adsorption layers was explained with the combined effect of several strong interactions, including H-bonds, intermediate dipole-dipole attraction, and long-range van der Waals attraction between the adsorbed molecules.

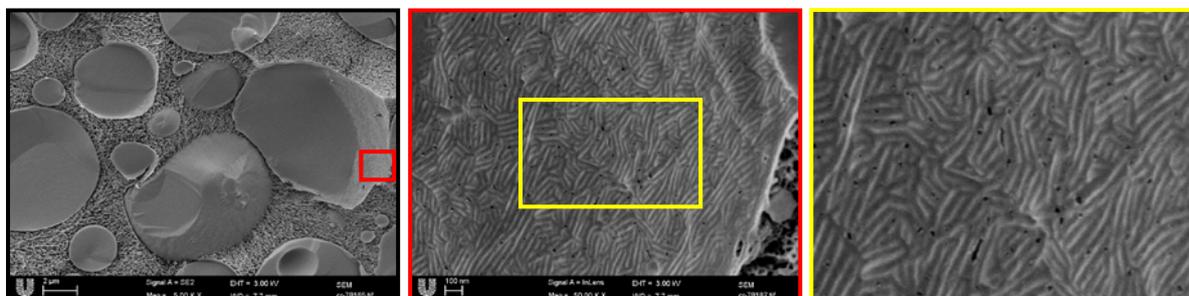

**Figure 4.** Freeze fracture SEM pictures of 65wt% sunflower oil emulsion stabilized with 1wt% solution of FD.

## 3.4. Origin of the extremely high elasticity of emulsions, stabilized by *Yucca Schidigera* saponins.

From the experimental results described so far we can conclude that the high shear elasticity of Yucca-stabilized emulsions, prepared with non-purified oil, is related to the presence of phytosterols in the oily phase which interact with the saponins from the aqueous phase and form elastic skin on the drop surface. However, as showed in our previous study (Tsibranska et al., 2020) the high interfacial elasticity (*per se*) can increase the shear elasticity of bulk emulsion by up to 4 times, as compared to the elasticity of emulsions with fluid adsorption layers. On the other hand, in the current study we found that the emulsion elasticity increased by ca. 20 times after changing the purified with non-purified oil. Therefore, there should be some additional effect to explain the observed extra high elasticity of these emulsion. To reveal the origin of this additional increase of emulsion elasticity, we performed microscope observations of the droplets in the emulsions prepared with non-purified and purified oil, see Figure 5.

From the images shown in Figure 5 one sees that drops of irregular shape are often seen in the emulsions prepared with FD and non-purified SFO, which is related to the skin formation on the drop surface when the Yucca saponins are combined with the phytosterols from the non-purified SFO. After oil purification, the oil drops are spherical due to the fluidization of the saponin adsorption layer. The second very important observation for these emulsions is that the oil drops are highly flocculated in all these emulsions. The flocculation is very pronounced not only for the emulsions prepared with non-purified SFO, but also for the emulsions prepared with purified SFO. For comparison, such flocculation is not observed for the hexadecane emulsions stabilized by QD (Figure 5D), for which the high dilatational elasticity leads to ca. 4 times increase of the emulsion elasticity in comparison to the conventional emulsions with low surface elasticity. All these results lead to the conclusion that the extra high shear elasticity of the *Yucca*-stabilized emulsions results from the combined action of the skin formation on the drop surface and the strong drop-drop attraction in these emulsions.

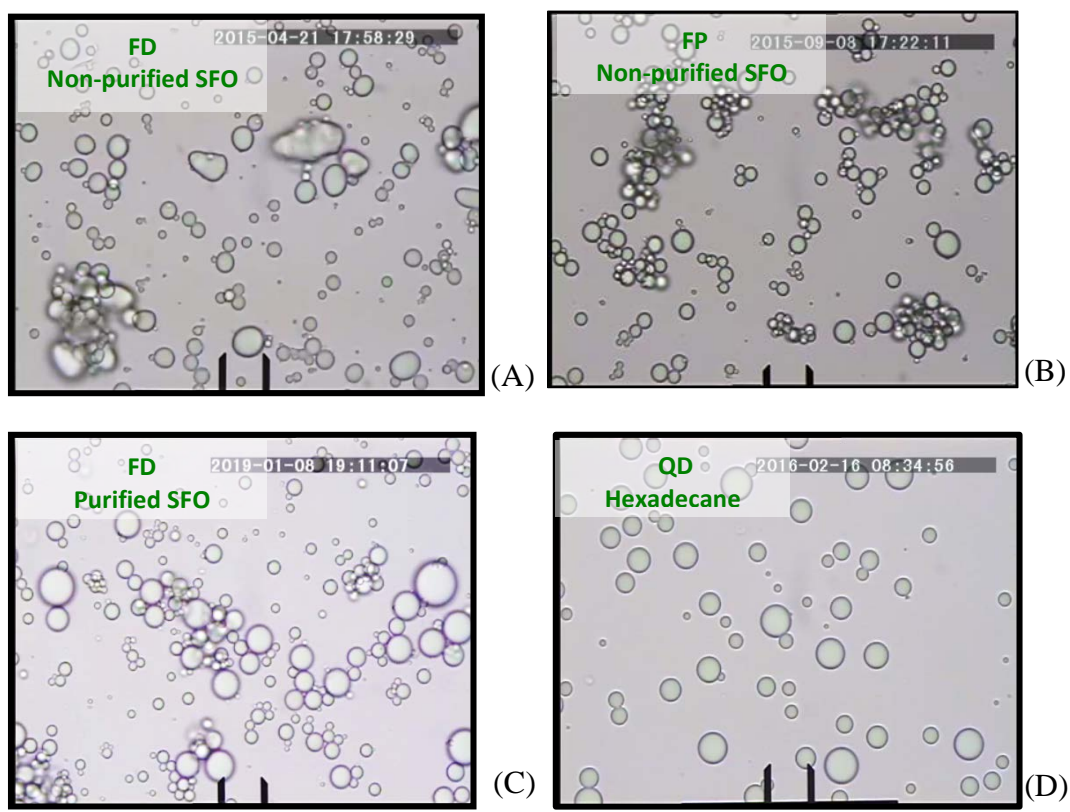

**Figure 5**. Images of emulsion drops of (A,B) non-purified sunflower oil and (C) purified SFO, stabilized with extracts (A,C) Foamation Dry 50; (B) Foamation P; (D) QD + non-purified hexadecane drops in the last emulsion. $\Phi_{SFO}$ = 75 wt. %. Emulsifier concentration in the aqueous phase: 1 wt. % saponin or 5 wt. % ROX. Distance between the scale bars: 20 μm. The drops are diluted in the serum taken from the same emulsion after centrifugation.

To confirm the hypothesis that a strong attraction between the drop surfaces leads to the observed drop flocculation in the *Yucca*-stabilized emulsions, we performed model experiments with emulsion films. Figure 6 presents microscope images of thin emulsion films stabilized by FD and observed in reflected white light. From these images we can determine the equilibrium thickness of the final emulsion films to be ≈ 5 nm which corresponds to very thin Newton black films (NBF). The final emulsion film has homogeneous thickness and does not contain any trapped entities (e.g. polymeric molecules). No visible Newton rings can be seen at the periphery of the emulsion films which is an evidence for a large contact angle film-meniscus. Both the small film thickness and the large contact angle are direct evidences for very strong attraction (adhesion) between the film surfaces, in agreement with the drop aggregation observed via optical microscopy (Kralchevsky, Danov, & Denkov, 2008). This attraction should of depletion type, not bridging by polymers, as the latter leads to non-homogeneous in thickness films (Dimitrova& Leal-Calderon 2004).

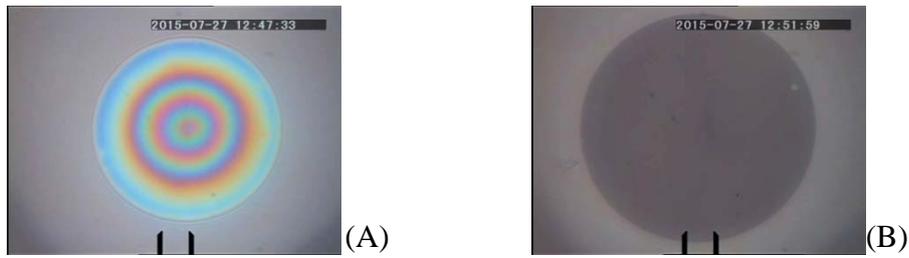

**Figure 6.** Emulsion films stabilized with Foamation Dry 50. (A) Immediately after film formation when a thicker dimple is present in the film center; (B) 4.5 minutes after film formation, when an ultrathin equilibrium Newton black film is finally reached.

To obtain more information about the role of drop-drop interactions and about the structure of the respective emulsions, we observed the kinetics of creaming of diluted emulsion with oil weight fraction of 9 wt. %, see section 2.6. These emulsions were placed in a glass cylinder and the change of the height of the cream with time, $h(t)$, was monitored. In Figure S3A we present results for the scaled cream height, $h(t)/h_0$, vs. time for FD-stabilized emulsions. One sees 3 distinct stages of the cream evolution. Initially $h$ does not change with time (stage 1) for a given period which we call "delay time", $t_D$. Afterwards, the cream starts compacting rapidly (stage 2) until $h$ reaches a steady-state value, and does not change further with time (stage 3). This particular creaming dynamics can be explained in the following manner. Initially the aggregated drops form a three-dimensional network in the emulsion, which spans the entire continuous phase. This structure is supported by the physical attraction between the neighboring drops, see Figure S3B-1. Under gravity, the diluted emulsions start compacting at an initially slow rate, which increases with time, see Figure S3B-2. At a certain moment, the cream cannot be compacted further and its height remains constant, see Figure S3B-3. Similar dynamics was observed with emulsions stabilized by the other two *Yucca Schidigera* extracts, F50 and FP. The delay time was $t_D \approx 100$ s for FD and FP, and $t_D \approx 60$ s for F50. The final cream which remained stable in these experiments had oil weight fraction of ca. 50 wt%.

Several publications (Blijdenstein, der Linden, van Vliet, & van Aken, 2004; Parker & Gunning, 1995; Manoj, Fillery-Travis, Watson, Hibberd, & Robins, 1998) have shown that such type of creaming is characteristic for emulsion drops which are aggregated due to depletion attraction.

We remind also that the emulsions of non-purified SFO with intermediate oil fraction stabilized with FD were very stable to syneresis. No drainage was observed for more than 30 days at oil weight fraction of 65 wt. % and 67 vol. %. This unusually high emulsion stability to syneresis can be also explained with the formation of stable three-dimensional network of drops which spans the continuous phase and prevents the phase separation.

The creaming stability of diluted emulsions prepared with purified and with non-purified SFO is compared in Figure S3C. The creaming is much faster for the emulsions prepared with purified SFO, but the creaming stages are the same. Therefore, for both types of emulsions, prepared with purified and non-purified oil, we have depletion attraction between the drops as confirmed in the microscope observations with emulsion films. On the other hand, the delay time is much longer for the emulsions prepared with non-purified oil, because the surface elasticity is high and the drops cannot rearrange easily with respect to each other in these emulsions. Hence, the compaction is much slower. In contrast, when purified SFO is used for emulsion preparation, the drops surfaces are tangentiality mobile, the drops can rearrange with respect to each other, and the compacting is faster.

### 3.5. Relative importance of the factors affecting the elasticity of bulk emulsions.

It is now well recognized that the elasticity of concentrated emulsions and foams is determined by several types of factors, including the volume fraction of the dispersed phase, $\Phi_V$, interfacial tension, $\sigma$, and average volume-surface radius, $R_{32}$, of the fluid particles – drops or bubbles (Princen, 2001; Denkov, Tcholakova, Hoehler, & Cohen-Addad, 2012; Mason et al., 1997). The effect of these factors is relatively well understood and described by theoretical expressions, separating the effect of the average capillary pressure, $\sigma/R_{32}$, from the effect of particle volume fraction, $\Phi_V$. Princen (Princen, 2001) and Mason et al., 1997, proposed empirical equations to describe the dependence of the dimensionless elastic modulus, $G'R_{32}/\sigma$ (scaled by the average capillary pressure, $\sigma/R_{32}$) on $\Phi_V$. Below we use as a reference the expression proposed by Mason:

$$\tilde{G}` = \frac{G`R_{32}}{\sigma} = 1.7\Phi_V(\Phi_V - \Phi_0); \qquad \Phi_0 = 0.64$$

The predictions of Princen, 2001 model are numerically different, but the conclusions from our study remain the same if we use the latter model.

It has been shown by several research groups that two additional factors can also affect strongly the elasticity of concentrated emulsions and foams: (1) The interfacial elasticity of the adsorption layers formed on the drop/bubble surfaces (Dimitrova & Leal-Calderon, 2001; Dimitrova & Leal-Calderon, 2004; Arditty, Schmitt, Giermanska-Kahn, & Leal-Calderon, 2004; Bressy, Hébraud, Schmitt, & Bibette, 2003; Tsibranska et al. 2020); and (2) The strength of the adhesion between the neighboring drops (Alben, Holtze, Tadros, & Schurtenberger, 2012; Becu, Manneville, & Colin, 2006) or bubbles (Denkov, Tcholakova, Hoehler, & Cohen-Addad, 2012; Princen, 2001; Politova et al., 2012). Below we analyze the available experimental results, both ours and of other authors, to clarify the relative importance of these two effects and of their combined action.

With this aim in view, in Table S1 in Supplementary material we compare the experimental results obtained with the three types of emulsions studied in literature: (A) regular emulsions with low interfacial elasticity and weak drop-drop adhesion; (B) with low interfacial elasticity and strong drop-drop adhesion, (C) with high interfacial elasticity and weak drop-drop attraction, and (D) with high interfacial elasticity and strong drop-drop adhesion. This comparison should be made for emulsions with similar oil volume fractions, in terms of the dimensionless elastic modulus to separate the known effects of $\sigma/R_{32}$ and $\Phi_V$.

Table S1 presents the measured bulk elastic moduli of various emulsions, stabilized by triterpenoid saponins (QD, Escin an Tea saponin), steroid saponin (FD), milk proteins, and synthetic surfactants (SDS, ROX). We compare the results with those obtained with conventional emulsions (low surface elasticity, negligible drop-drop adhesion) and with the prediction by the model of Mason et al., 1997. All compared emulsions have very similar drop volume fractions, $\Phi_V \approx 75$ vol. %, for which the model of Mason et al., 1997, predicts dimensionless emulsion elastic modulus, $\widetilde{G}' \approx 0.15$. The results shown in Table S1 confirm that both the interfacial elasticity and the drop-drop attraction increase the elasticity of bulk emulsions up to several times when they are present separately. Remarkably, the emulsions in which both factors are combined show synergistically higher bulk elasticity which could exceed the theoretical predictions by two orders of magnitude, as it is the case with the emulsions stabilized with FD, for which $\widetilde{G}' \approx 9.6$.

### 4. Main results and conclusions.

The measured extremely high elasticity of bulk emulsions, prepared with non-purified SFO and saponin extract of *Yucca Schidigera* plant, is due to strong interactions between the phytosterols present in the non-purified oil and the saponins in the extract. These interactions resulted in the formation of highly elastic mixed adsorption layer on the surface of the oil drops. This elastic layer, combined with strong attraction between the neighboring emulsion drops, leads to synergistic effect with resulting extremely high emulsion elasticity. The drop-drop adhesion is due to depletion interactions induced by some of the component(s), polymer or surfactant aggregates, present in the plant extract.

Interestingly, after purification of the SFO which removes the phytosterols, the adhesion between the drops remains still very strong, while the interfacial elasticity becomes very low and, as a consequence, the emulsion elasticity decreases by > 50 times. This specific interaction between the phytosterols and Yucca saponins can be probably used in other systems to boost the interfacial elasticity and the bulk elasticity of the respective emulsions.

The comparison of the experimental results in the current study with results from previous studies shows that in all systems, in which both the high interfacial elasticity and the drop-drop adhesion are combined, the elasticity of the respective emulsions is much higher as compared to the systems in which only one of these two effects is present. Therefore, the observed synergistic effect between the high interfacial elasticity and drop-drop adhesion is rather general and can be used as a universal approach to produce highly elastic emulsions and foams.

**Acknowledgements**

The authors are grateful to C. Remijn, Unilever R&D, Vlaardingen, The Netherlands for performing cryo-SEM observations.

**Funding:** This work was partially supported by Unilever R&D Vlaardingen, the Netherlands and by the Bulgarian Ministry of Education and Science under the National Research Programme "Healthy Foods for a Strong Bio-Economy and Quality of Life" approved by DCM # 577/17.08.2018. S. Tsibranska is grateful to Operational program "Science and Education for Smart Growth", project BG05M2OP001-2.009-0028 for the financial support for scientific visit at Unilever R&D Vlaardingen, the Netherlands.

**Conflict of interest:** None

# Supplementary materials

## Origin of the extremely high elasticity of bulk emulsions, stabilized by *Yucca Schidigera* saponins

Sonya TSIBRANSKA,[1] Slavka TCHOLAKOVA,[1*] Konstantin GOLEMANOV,[1] Nikolai DENKOV,[1] Luben ARNAUDOV,[2] Eddie PELAN,[2] Simeon D. STOYANOV[2,3,4]

[1]*Department of Chemical and Pharmaceutical Engineering*
*Faculty of Chemistry and Pharmacy, Sofia University*
*1 J. Bourchier Ave., 1164 Sofia, Bulgaria*

[2] *Unilever R&D, Vlaardingen, The Netherlands*

[3]*Laboratory of Physical Chemistry and Colloid Science, Wageningen University, 6703 HB Wageningen, The Netherlands*

[4]*Department of Mechanical Engineering, University College London, Torrington Place, London WC1E 7JE, UK*

*Corresponding author:
Prof. Slavka Tcholakova
Department of Chemical and Pharmaceutical Engineering
Faculty of Chemistry and Pharmacy, Sofia University
1 James Bourchier Ave., 1164 Sofia
Bulgaria

Phone: (+359-2) 962 5310
Fax:    (+359-2) 962 5643
E-mail: SC@LCPE.UNI-SOFIA.BG

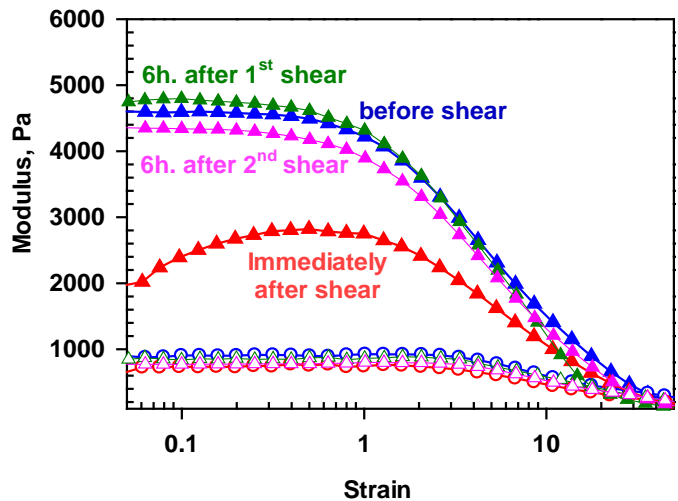

**Figure S1.** Regeneration of the viscoelasticity of emulsions of sunflower oil-in-water, stabilized with FD after pre-shear. Elastic (full symbols) and viscous (empty symbols) as a function of amplitude of deformation at 1 Hz;

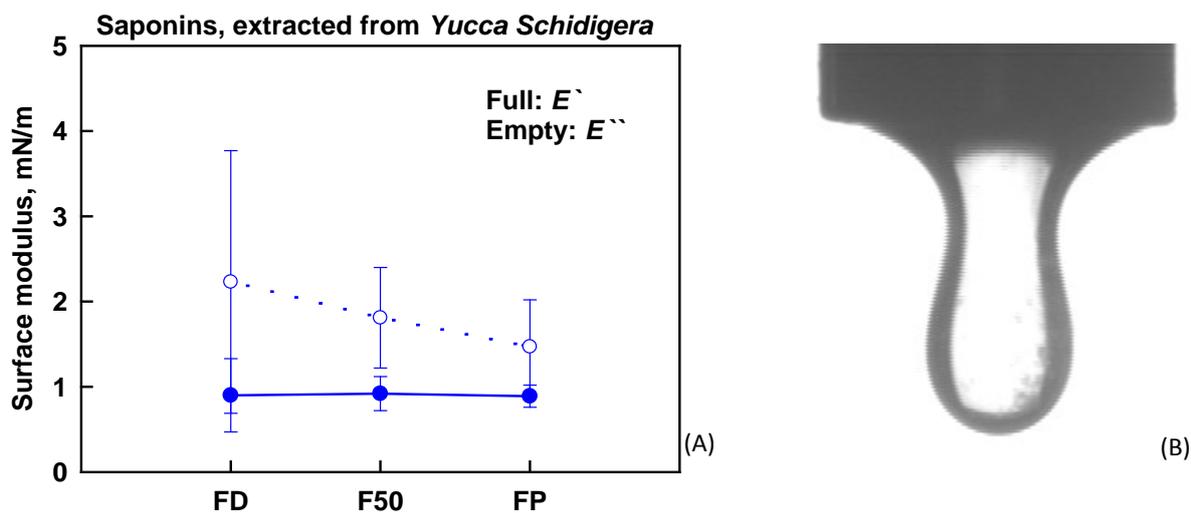

**Figure S2.** (A) Surface modulus at 0.8-1.2% deformation as a function of Yucca extracts. Saponin concentration 1 wt % FD in purified oil; (B) Drop of 1 wt % solution of FD in sunflower oil + 0.15 wt % DG. No wrinkles are observed in that case.

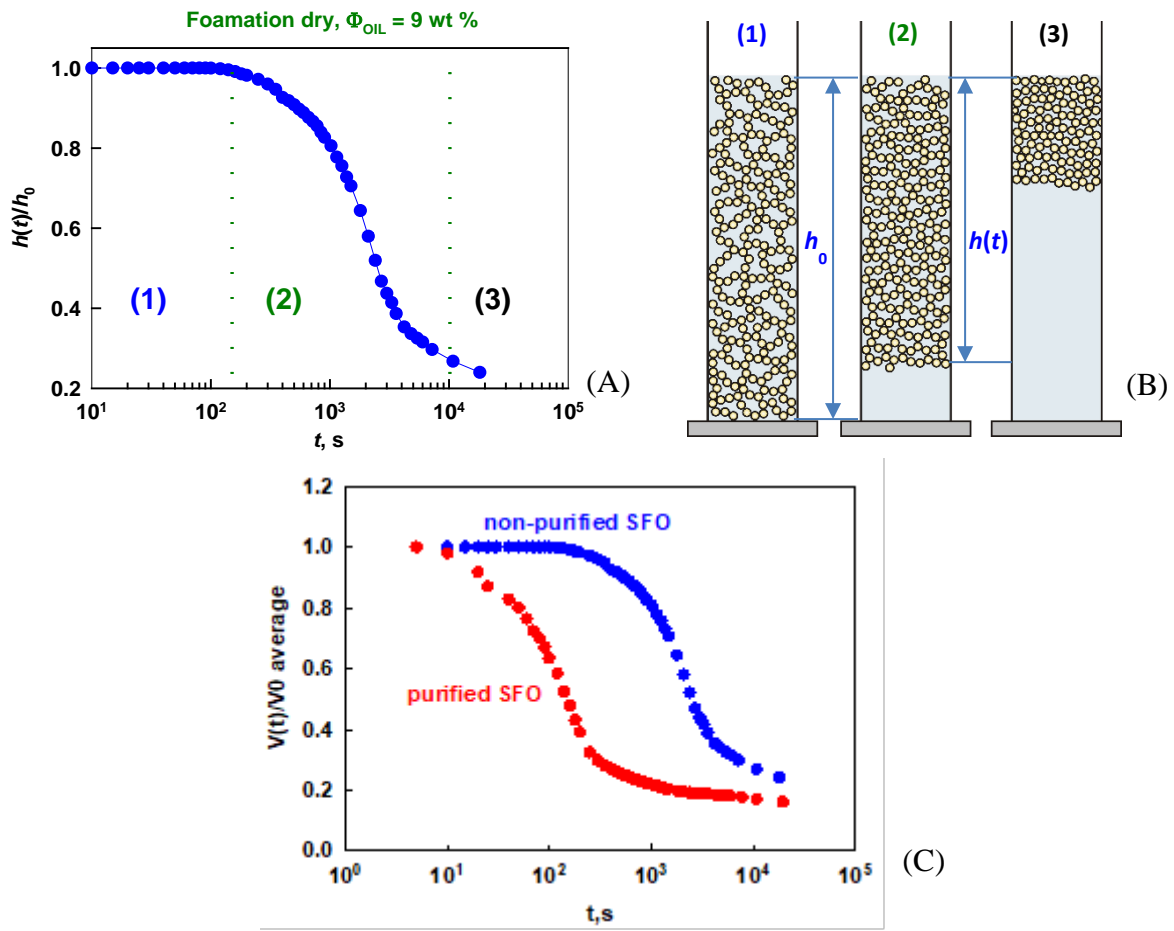

**Figure S3.** Dynamics of creaming of emulsions, stabilized with solution of FD. (A) $h(t)/h_0$ vs. time; (B) Illustrative sketches of the stages of the creaming process. Concentrations: 1 wt % FD in the aqueous phase; $\Phi_{SFO}$ = 9 wt. %; (C) Comparison between the creaming dynamics of FD-stabilized emulsion with purified and non-purified SFO

**Table S1**. Dimensionless elastic modulus, G'$R_{32}$/σ ($R_{32}$ is a mean volume surface radius and σ is the interfacial tension), of regular emulsions (low elasticity and weak drop-drop adhesion); emulsions with strong drop adhesion (no interfacial elasticity); emulsions with high interfacial elasticity (no drop adhesion); and emulsions with high interfacial elasticity and drop adhesion.

| Type of system | Emulsifier | Oil | $\Phi_V$, % | Dimensionless elastic modulus | Reference |
|---|---|---|---|---|---|
| Prediction of Mason et al. | - | - | 75 | 0.15 | Mason et al. 1996 |
|  | - | - | 80 | 0.22 |  |
| Regular emulsions (Type A) | SDS | SFO | 75 | 0.17 | Tsibranska et al. 2019 |
|  | SDS | Hexadecane |  | 0.10 |  |
|  | Tea saponin | SFO |  | 0.14 |  |
| Strong drop-drop adhesion (Type B) | ROX | SFO |  | 0.22 |  |
|  | Tea saponin | Hexadecane |  | 0.23 |  |
| High interfacial elasticity (Type C) | QD | Hexadecane |  | 0.53 |  |
|  | Escin | SFO |  | 0.34 |  |
| High interfacial elasticity + Strong drop-drop adhesion (Type D) | β-lactoglobulin | Hexadecane | 75 | ≈ 0.14 ÷ 0.20 | Dimitrova, & Leal-Calderon 2004 |
|  | Lyzozyme | Hexadecane |  | ≈ 0.30 ÷ 0.40 |  |
|  | Bovine serum albumin | Hexadecane |  | ≈ 0.66 |  |
|  | Sodium caseinate | SBO |  | ≈ 6 | Bressy et al. 2003 |
|  | FD | SFO | 76.5 | 9.6 | Current work |

26